\documentclass{optica-article}

\journal{opticajournal} 

\articletype{Research Article}

\usepackage{amsmath}
\graphicspath{{figure/}{./figure/}}

\begin{document}

\title{Nonreciprocal state-selective splitting of eigenstates at a polarization exceptional point}

\author{Zhuofan Yu,\authormark{1,\dag} Soojeong Baek,\authormark{2,\dag} Puxuan Jiang,\authormark{1} Junqing Li,\authormark{1} Teun-Teun Kim,\authormark{2,*} Xingguang Liu,\authormark{1,3,*} and Yongkang Dong\authormark{4,5}}
\address{\authormark{1}School of Physics, Harbin Institute of Technology, Harbin 150001, China\\
\authormark{2}Department of Physics and Energy Harvest Storage Research Center, University of Ulsan, 93 Daehak-ro, Nam-gu, Ulsan 44610, Republic of Korea\\
\authormark{3}Heilongjiang Provincial Key Laboratory of Advanced Quantum Functional Materials and Sensor Devices, Harbin 150001, China\\
\authormark{4}National Key Laboratory of Laser Spatial Information, Harbin Institute of Technology, Harbin 150001, China\\
\authormark{5}Zhengzhou Research Institute, Harbin Institute of Technology, Zhengzhou 450007, China\\
\authormark{\dag}These authors contributed equally to this work.}
\email{\authormark{*}ttkim@ulsan.ac.kr, Liu\_xg@hit.edu.cn}

\begin{abstract*}
Non-Hermitian photonic systems can be engineered to exhibit exceptional points (EPs), where eigenvalues and eigenstates coalesce. Although perturbations near EPs have mainly been characterized through eigenvalue splitting, the branch-resolved evolution of the coalesced eigenstates under a tunable reciprocity-breaking perturbation remains less systematically explored. Here, we show that tunable nonreciprocal coupling can produce state-selective splitting of the coalesced eigenstates near a second-order polarization EP. Using temporal coupled-mode theory, we find that the corresponding off-diagonal perturbation produces a triangular Hamiltonian in the circular-polarization basis. In the ideal limit, one eigenstate remains pinned to the EP state, whereas the other is continuously reoriented by the perturbation. In a realistic scattering system, residual reverse-polarization conversion weakly displaces the nominally pinned branch, resulting in quasi-pinning. We numerically demonstrate this mechanism in a reflective magneto-optical metasurface supporting a circular-polarization EP of the reflection matrix near 582~nm. Full-wave simulations show that a perpendicular magneto-optical bias leaves one reflection eigenstate close to the EP state while reversibly displacing the other on the Poincar\'e sphere. This asymmetric eigenstate evolution manifests as a magnetically tunable circular-dichroic reflection response. Our results establish a nonreciprocal route to state-selective eigenstate control near polarization EPs.
\end{abstract*}

\section{Introduction}
Non-Hermitian photonics provides a framework for describing open optical systems whose effective Hamiltonians are governed by loss, gain, radiation, and engineered coupling \cite{elganainy2018NonHermitianPhysics,miri2019ExceptionalPointsOptics,ozdemir2019ParitytimeSymmetryExceptional,meng2024exceptional}. A defining feature of such systems is the exceptional point (EP), a spectral degeneracy at which both the eigenvalues and eigenstates coalesce \cite{heiss2012PhysicsExceptionalPoints,berry2004PhysicsNonhermitianDegeneracies,qin2026PhotonicExceptionalPoints,lee2023chiral}. The resulting square-root eigenvalue topology has enabled unusual responses in sensing \cite{chen2017ExceptionalPointsEnhance,hodaei2017EnhancedSensitivityHigher}, lasing \cite{brandstetter2014ReversingPumpDependence}, and mode conversion \cite{doppler2016DynamicallyEncirclingExceptional,nasari2022ObservationChiralState,yoon2018TimeAsymmetricLoop}. Most studies of perturbations near EPs have consequently focused on how the coalesced eigenvalues separate. The accompanying evolution of the eigenstates, however, is equally important because it determines how the modal or polarization content of the system can be controlled after the degeneracy is lifted. In particular, the branch-resolved evolution of EP eigenstates under a tunable reciprocity-breaking perturbation remains less systematically explored.

Metasurfaces provide a versatile platform for investigating polarization EPs because their subwavelength meta-atoms enable simultaneous control over phase, amplitude, loss, and polarization conversion \cite{yu2014FlatOpticsDesigner,yu2011LightPropagationPhase,kildishev2013PlanarPhotonicsMetasurfaces}. Polarization EPs have been demonstrated in plasmonic structures, dielectric metasurfaces, and photonic-crystal slabs through loss engineering, geometric asymmetry, and resonant-mode coupling \cite{kodigala2016ExceptionalPointsThree,song2021PlasmonicTopologicalMetasurface,he2023ScatteringExceptionalPoint,shou2025ResonantScatteringExceptional,dong2020LossAssistedMetasurface,zhen2015SpawningRingsExceptional}, leading to chiral absorption \cite{sweeney2019PerfectlyAbsorbingExceptional,park2020ObservationExceptionalPoint,kang2016ChiralExceptionalPoints}, polarization-state control \cite{yang2024CreatingPairsExceptional,qin2025SphereArbitrarilyPolarized,wu2024TwinsExceptionalPoints}, and parity--time-symmetry breaking \cite{lawrence2014ManifestationPT}. Actively tunable EP platforms have also been developed using electrically or optically controlled material responses~\cite{ergoktas2022TopologicalEngineeringTerahertz,baek2023NonHermitianChiral,ding2024ElectricallyTunableTopological,wang2025PhotoswitchableExceptionalPoints}. Such tuning has primarily been used to access, shift, or traverse the degeneracy and to probe its eigenvalue and eigenstate topology. By comparison, active control over the unequal evolution of the two eigenstate branches after the EP is lifted remains less developed.

For a reciprocal perturbation that modifies the two coupling directions symmetrically, the two eigenstates generally emerge from the coalesced state as paired branches. A qualitatively different response becomes possible when the perturbation distinguishes the two coupling directions and preserves an invariant polarization subspace. In an appropriate basis, this condition appears as a triangular effective matrix, allowing one eigenstate to remain fixed while the other evolves with the perturbation. Related block-triangular non-Hermitian structures have recently been investigated in the context of exceptional deficiency, where entire eigenspaces coalesce and exhibit selective dynamics in active mechanical lattices \cite{li2026exceptional,ryu2024ExceptionalClassifications}. Here, we consider the complementary two-state problem of perturbatively lifting an isolated second-order polarization EP and selectively controlling the resulting eigenstate branches.

The magneto-optical effect provides a physical route to such branch-selective control. Under a static magnetic bias, a gyrotropic medium acquires antisymmetric off-diagonal permittivity components, breaking time-reversal symmetry and producing nonreciprocal polarization coupling \cite{sounas2017NonReciprocalPhotonics,bi2011ChipOpticalIsolation,khanikaev2010OneWayExtraordinary,fan2019MagnetoopticalMetamaterialsNonreciprocal,belotelov2011EnhancedMagnetoOptical,belotelov2013PlasmonMediatedMagneto}. Unlike a reciprocal perturbation, this coupling modifies the two directional conversion pathways with opposite signs. Its magnitude and sign can therefore be used to control not only the splitting of the EP but also the direction and extent of the resulting eigenstate evolution. A magneto-optical metasurface is particularly suitable for examining this response because its reflection eigenstates can be directly represented and tracked on the Poincar\'e sphere.

In this work, we theoretically identify and numerically demonstrate state-selective eigenstate splitting induced by antisymmetric nonreciprocal coupling near a second-order polarization EP. Using temporal coupled-mode theory, we show that the Hamiltonian becomes triangular in the circular-polarization basis at the chosen chiral EP. Consequently, one eigenstate remains pinned to the EP polarization, whereas the other is continuously reoriented by the nonreciprocal perturbation. We then extend the analysis to a realistic scattering matrix and derive a robustness criterion for quasi-pinning in the presence of residual reverse-polarization conversion. As a representative implementation, we design a reflective magneto-optical metasurface supporting a circular-polarization EP of the reflection matrix near 582~nm. Full-wave simulations show that varying the magneto-optical bias leaves one reflection eigenstate close to the EP state while reversibly displacing the other on the Poincar\'e sphere. This branch-selective eigenstate evolution manifests as a continuously tunable circular-dichroic reflection response.

\section{State-selective eigenstate splitting at a polarization EP}
\subsection{Minimal model for nonreciprocal perturbations around an EP}
\label{sec:model}
To isolate how reciprocity breaking modifies the two eigenstate branches emerging from an EP, we consider two orthogonally polarized resonances described by the minimal non-Hermitian Hamiltonian
\begin{equation}
  H_{\mathrm{nr}}=
  \begin{pmatrix}
    \omega_{x}+i\gamma_{x} & \kappa+i\beta\tilde{g}_{B}\\[2pt]
    \kappa-i\beta\tilde{g}_{B} & \omega_{y}+i\gamma_{y}
  \end{pmatrix},
  \label{eq:Hnr}
\end{equation}
where $\omega_{j}+i\gamma_{j}$ denotes the complex frequency of mode $j=x,y$. We adopt the $e^{i\omega t}$ convention, such that $\gamma_{j}>0$ represents modal loss. In the absence of time-reversal-symmetry breaking, the two modes are reciprocally coupled through the equal off-diagonal elements $\kappa$. The magneto-optical perturbation adds equal-magnitude corrections with opposite signs to the two polarization-conversion channels. Here, $\tilde{g}_{B}$ is treated as a general parameter that sets the strength and sign of the nonreciprocal coupling perturbation, while $\beta$ converts it to the modal-coupling scale, so that $\beta\tilde{g}_{B}$, like $\omega_{j}$, $\gamma_{j}$, and $\kappa$, has units of angular frequency; the units of $\beta$ therefore follow from those of $\tilde{g}_{B}$. In the metasurface considered below, this term originates from the antisymmetric off-diagonal components of the gyrotropic permittivity tensor, where $\tilde{g}_{B}$ becomes the dimensionless magneto-optical parameter of Eq.~(\ref{eq:eps}) and $\beta$ accordingly carries units of angular frequency.

\begin{figure}[!htb]
\centering\includegraphics[width=\linewidth]{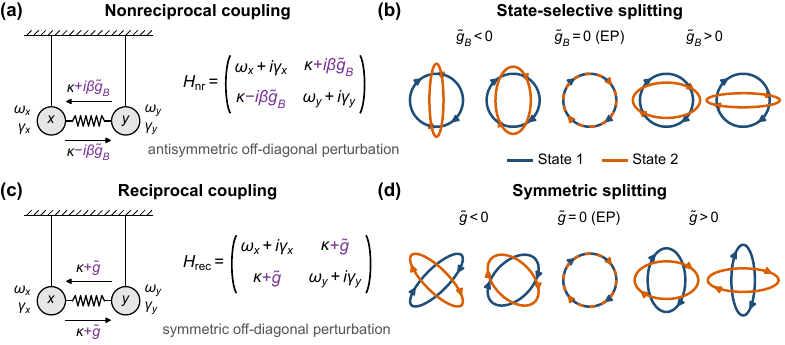}
\caption{
\textbf{Comparison of the effects of nonreciprocal and reciprocal perturbations on the EP eigenstates. Nonreciprocal perturbations lead to state-selective EP splitting, whereas reciprocal perturbations produce symmetric splitting of the two eigenstates.}
\textbf{(a)} Left: coupled-oscillator schematic of the two orthogonally polarized modes $x$ and $y$, with complex frequencies $\omega_{x,y}+i\gamma_{x,y}$ and coupling $\kappa$; the arrows denote the two coupling directions, which acquire opposite corrections $\pm i\beta\tilde{g}_B$. Right: nonreciprocal-perturbation Hamiltonian $H_\mathrm{nr}$, where the nonreciprocal term $\pm i\beta\tilde{g}_B$ enters the coupling matrix as an antisymmetric off-diagonal element.
\textbf{(b)} Eigen-polarization evolution under a nonreciprocal perturbation (state-selective splitting). At $\tilde{g}_B = 0$ the system is at the EP and the two eigenstates merge into the same circular-polarization state; as $\tilde{g}_B$ departs from zero, State~1 (blue) is pinned near the EP circular-polarization state, while State~2 (orange) evolves into an elliptical polarization state with the opposite major-axis orientation.
\textbf{(c)} Left: the same coupled-oscillator schematic under a reciprocal perturbation, in which both coupling directions acquire the same correction $\tilde{g}$. Right: reciprocal-perturbation Hamiltonian $H_\mathrm{rec}$, where the symmetric coupling perturbation $\tilde{g}$ appears identically in the two off-diagonal elements.
\textbf{(d)} Under a reciprocal perturbation, the two eigenstates leave the EP simultaneously with equal magnitude and orthogonal major axes, exhibiting symmetric splitting.}
\label{fig:hamiltonian_compare}
\end{figure}

\begin{figure}[!htb]
\centering\includegraphics[width=\linewidth]{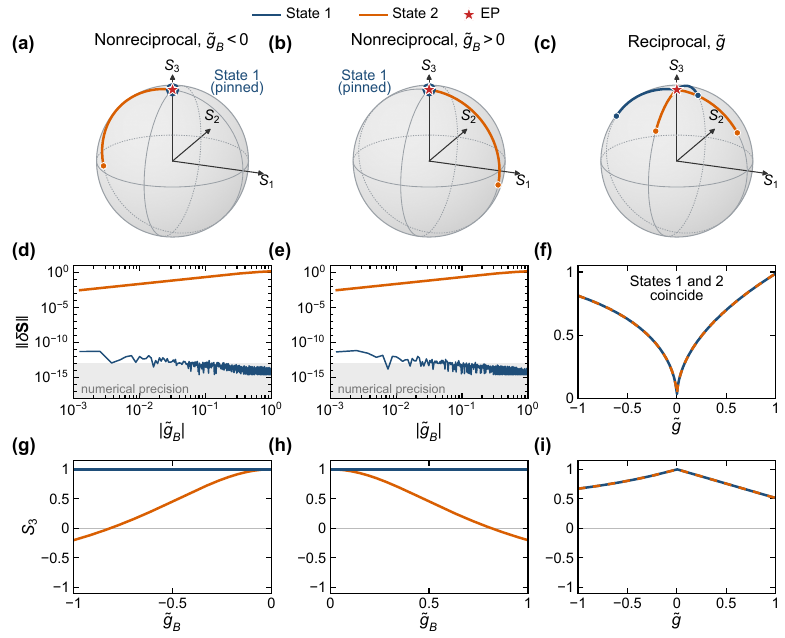}
\caption{
\textbf{Comparison of the effects of a nonreciprocal perturbation ($\tilde{g}_B$) and a reciprocal perturbation ($\tilde{g}$) on the eigenstates.}
\textbf{(a)--(c)} Trajectories of the two eigenstates on the Poincar\'e sphere; the star marks the EP.
\textbf{(a)} For $\tilde{g}_B<0$, State~1 (blue) is pinned at the EP while State~2 (orange) moves away from the EP along the sphere.
\textbf{(b)} For $\tilde{g}_B>0$, the evolution of State~2 is symmetric to the negative-value case about the $S_3$ axis.
\textbf{(c)} Under a reciprocal perturbation $\tilde{g}$, the two eigenstates leave the EP symmetrically.
\textbf{(d)--(f)} Deviation of the eigenstates from the EP,
$\|\delta\mathbf{S}\|$
as a function of the perturbation.
\textbf{(d), (e)} State~1 (blue) stays at the numerical-precision limit (gray shaded region, about $10^{-13}$); State~2 (orange) increases with $|\tilde{g}_B|$.
\textbf{(f)} Under a reciprocal perturbation, the two eigenstates have equal distances from the EP.
\textbf{(g)--(i)} Evolution of the Stokes parameter $S_3$ (circular-polarization component) with the perturbation.
\textbf{(g), (h)} State~1 keeps $S_3 = 1$ (LCP at the EP); State~2 evolves away from LCP toward RCP, and the two panels are mirror images about $\tilde{g}_B = 0$.
\textbf{(i)} Under a reciprocal perturbation, the two eigenstates leave the EP symmetrically.}
\label{fig:comparison}
\end{figure}

At $\tilde{g}_{B}=0$, the unperturbed parameters are chosen to satisfy the second-order chiral-EP condition derived in Section~\ref{sec:analytic}. We focus on the EP whose coalesced eigenstate $|+\rangle$ corresponds to LCP under the polarization convention adopted here. The opposite-handed EP, whose coalesced eigenstate $|-\rangle$ corresponds to RCP, follows from the other branch of the EP condition. In Figs.~\ref{fig:hamiltonian_compare} and \ref{fig:comparison}, $\omega_{x}+i\gamma_{x}$, $\omega_{y}+i\gamma_{y}$, and $\kappa$ are held fixed at the unperturbed EP values, and only the perturbation parameter is varied; in these figures, $\tilde{g}_{B}$ and $\tilde{g}$ are normalized by their maximum values in the scan, so that they range from $-1$ to $1$. Thus, $\tilde{g}_{B}=0$ marks the EP, whereas finite $\tilde{g}_{B}$ lifts the degeneracy.%

In the ideal nonreciprocal model, the two eigenstate branches evolve unequally after the EP is lifted. One branch remains exactly fixed at the EP polarization, whereas the other is continuously reoriented according to the magnitude and sign of $\tilde{g}_{B}$. We refer to this unequal branch evolution as state-selective splitting. Hereafter, State~1 denotes the pinned branch and State~2 denotes the tunable branch.

For comparison, we consider a representative reciprocal perturbation that modifies the two coupling channels equally,
\begin{equation}
  H_{\mathrm{rec}}=
  \begin{pmatrix}
    \omega_{x}+i\gamma_{x} & \kappa+\tilde{g}\\[2pt]
    \kappa+\tilde{g} & \omega_{y}+i\gamma_{y}
  \end{pmatrix}.
  \label{eq:Hrec}
\end{equation}
Here, $\tilde{g}$ is a reciprocal coupling perturbation with the same units (angular frequency) as $\kappa$. At $\tilde{g}=0$, this model starts from the same unperturbed EP as $H_{\mathrm{nr}}$. For finite $\tilde{g}$, the two off-diagonal elements remain equal. This perturbation does not preserve an invariant eigenstate branch, and the two eigenstates depart from the EP in a paired manner.

Figure~\ref{fig:hamiltonian_compare} schematically illustrates how the symmetry of the off-diagonal perturbation determines the eigenstate evolution. In the nonreciprocal model, the perturbation enters the two coupling channels with opposite signs (Fig.~\ref{fig:hamiltonian_compare}(a)). Starting from the coalesced LCP state at $\tilde{g}_{B}=0$, State~1 remains fixed at the EP polarization, whereas State~2 evolves into an elliptically polarized state. Reversing the sign of $\tilde{g}_{B}$ reverses the orientation of the tunable polarization state while leaving the pinned branch unchanged (Fig.~\ref{fig:hamiltonian_compare}(b)). By contrast, the representative reciprocal perturbation modifies the two coupling channels equally (Fig.~\ref{fig:hamiltonian_compare}(c)). Both eigenstates then evolve away from the EP through paired polarization changes, with neither branch remaining fixed (Fig.~\ref{fig:hamiltonian_compare}(d)).

To quantify the displacement of each eigenstate from the EP polarization, we use the Euclidean distance between the corresponding normalized Stokes vectors,
\begin{equation}
  \lVert\delta\mathbf{S}\rVert=
  \sqrt{
    \left(S_{1}-S_{1,\mathrm{EP}}\right)^{2}
    +\left(S_{2}-S_{2,\mathrm{EP}}\right)^{2}
    +\left(S_{3}-S_{3,\mathrm{EP}}\right)^{2}
  }.
  \label{eq:dS}
\end{equation}
Figure~\ref{fig:comparison} provides a quantitative comparison of the three cases depicted schematically in Fig.~\ref{fig:hamiltonian_compare}. For $\tilde{g}_{B}<0$, State~1 remains exactly pinned at the EP polarization, whereas State~2 moves away along one side of the Poincar\'{e} sphere (Fig.~\ref{fig:comparison}(a)). The displacement of State~2 increases with $|\tilde{g}_{B}|$, while the small nonzero values of $\lVert\delta\mathbf{S}\rVert$ for State~1 remain at the numerical-precision floor (Fig.~\ref{fig:comparison}(d)). The corresponding change in $S_{3}$ shows the evolution of State~2 from LCP toward elliptically polarized states (Fig.~\ref{fig:comparison}(g)).

For $\tilde{g}_{B}>0$, State~1 remains pinned, whereas State~2 follows the opposite trajectory on the Poincar\'{e} sphere (Fig.~\ref{fig:comparison}(b)). The displacement magnitude and $S_{3}$ evolution are symmetric under reversal of $\tilde{g}_{B}$ (Figs.~\ref{fig:comparison}(e) and \ref{fig:comparison}(h)). Thus, $|\tilde{g}_{B}|$ controls the extent of the polarization displacement, while the sign of $\tilde{g}_{B}$ selects its direction.

For the representative reciprocal perturbation, both eigenstates depart from the EP along paired trajectories (Fig.~\ref{fig:comparison}(c)) and exhibit equal changes in $\lVert\delta\mathbf{S}\rVert$ and $S_{3}$ (Figs.~\ref{fig:comparison}(f) and \ref{fig:comparison}(i)). The contrast between these responses establishes the state-selective character of the nonreciprocal perturbation. We next derive the triangular matrix structure responsible for this exact pinning.

\subsection{Analytic origin of state-selective splitting}
\label{sec:analytic}
The state-selective response can be traced to the matrix structure obtained when the nonreciprocal perturbation is applied to the chiral EP. We define the mean frequency $\bar{\Omega}=(\omega_{x}+i\gamma_{x}+\omega_{y}+i\gamma_{y})/2$ and the half-difference $\Delta\Omega=(\omega_{x}+i\gamma_{x}-\omega_{y}-i\gamma_{y})/2$ of the complex modal frequencies, so that Eq.~\eqref{eq:Hnr} can be written as
\begin{equation}
  H_{\mathrm{nr}}=\bar{\Omega}\mathbf{\mathit{I}}+
  \begin{pmatrix}
    \Delta\Omega & \kappa+i\beta\tilde{g}_{B}\\[2pt]
    \kappa-i\beta\tilde{g}_{B} & -\Delta\Omega
  \end{pmatrix}.
  \label{eq:Hnr_split}
\end{equation}
At $\tilde{g}_{B}=0$, the two eigenvalues coalesce when $\left(\Delta\Omega\right)^{2}+\kappa^{2}=0$. We choose the chiral branch $\Delta\Omega=-i\kappa$, for which the coalesced eigenstate is $|\psi_{\mathrm{EP}}\rangle=|+\rangle=\left(1,\,i\right)^{\mathsf{T}}/\sqrt{2}$, corresponding to LCP under our convention. The opposite branch, $\Delta\Omega=+i\kappa$, produces the RCP counterpart with coalesced eigenstate $|-\rangle$.

At the LCP EP, transforming Eq.~\eqref{eq:Hnr_split} to the circular-polarization basis $\left(|+\rangle,|-\rangle\right)$ gives
\begin{equation}
  H_{\pm}=\bar{\Omega}\mathbf{\mathit{I}}+
  \begin{pmatrix}
    -\beta\tilde{g}_{B} & -2i\kappa\\[2pt]
    0 & \beta\tilde{g}_{B}
  \end{pmatrix}.
  \label{eq:Hpm}
\end{equation}
The vanishing lower-left matrix element defines an invariant one-dimensional subspace spanned by $|+\rangle$. Consequently, one eigenstate remains independent of the perturbation, $|\psi_{\mathrm{pin}}\rangle=|+\rangle$, whereas the other eigenstate, written up to normalization, is
\begin{equation}
  |\psi_{\mathrm{tun}}\rangle=
  |+\rangle+\frac{i\beta\tilde{g}_{B}}{\kappa}|-\rangle.
  \label{eq:psitun}
\end{equation}
Equation~\eqref{eq:psitun} shows that the magnitude of $\tilde{g}_{B}$ controls the admixture of the opposite circular component, while its sign reverses the orientation of the resulting polarization state. The tunable branch is therefore generally an elliptically polarized superposition rather than a pure state of the opposite handedness. In contrast, the pinned branch remains exactly at the original LCP EP state for any $\tilde{g}_{B}$ within the ideal model.

The difference from the reciprocal perturbation can be seen by transforming Eq.~\eqref{eq:Hrec} at the same unperturbed EP. In the circular basis,
\begin{equation}
  H_{\mathrm{rec},\pm}=\bar{\Omega}\mathbf{\mathit{I}}+
  \begin{pmatrix}
    0 & -i\left(2\kappa+\tilde{g}\right)\\[2pt]
    i\tilde{g} & 0
  \end{pmatrix}.
  \label{eq:Hrecpm}
\end{equation}
For $\tilde{g}\neq0$, both off-diagonal elements are finite. The $|+\rangle$ subspace is therefore no longer invariant, and neither eigenstate remains pinned. This direct comparison shows that state-selective splitting does not follow merely from lifting an EP; it results from the triangular structure created by the antisymmetric perturbation to the two coupling channels.

Exact pinning of $|+\rangle$ relies on the vanishing lower-left element of Eq.~\eqref{eq:Hpm}. A realistic scattering matrix generally retains a finite residual reverse-polarization-conversion channel. In Section~\ref{sec:robust}, we examine how this deviation from ideal triangularity converts exact pinning into quasi-pinning and derive the condition under which the state-selective response remains robust.

\subsection{Reflection-matrix manifestation and robustness of state-selective splitting}
\label{sec:robust}
For a physical implementation, the eigenstate mechanism derived from the modal Hamiltonian must be connected to the polarization-resolved scattering response. The effective Hamiltonian in Section~\ref{sec:analytic} governs the internal amplitudes of the two polarization resonances, whereas the experimentally accessible quantity is the reflection matrix that relates the incident and reflected polarization states. These matrices are distinct operators and need not share the same eigenvalues. Near an isolated two-mode resonance, however, temporal coupled-mode theory maps the internal modal dynamics onto the scattering response~\cite{fan2003TemporalCoupledMode}. As detailed in Supplementary Note~1, the circular-basis reflection matrix inherits an analogous triangular structure in the ideal limit, thereby preserving the pinned polarization eigenstate.

Near the LCP reflection-matrix EP, we write the local circular-basis response as
\begin{equation}
  R_{\mathrm{c}}=r_{0}\mathbf{\mathit{I}}+
  \begin{pmatrix}
    \Delta r_{B}/2 & b\\[2pt]
    \xi & -\Delta r_{B}/2
  \end{pmatrix},
  \label{eq:Rc_local}
\end{equation}
where $r_{0}=(r_{++}+r_{--})/2$ is the common (mean) co-polarized reflection coefficient, $\Delta r_{B}=r_{++}-r_{--}$ is the difference between the two co-polarized circular reflection channels, $b=r_{+-}$ is the polarization-conversion channel associated with the Jordan coupling, and $\xi=r_{-+}$ represents residual reverse-polarization conversion. The common term $r_{0}\mathbf{\mathit{I}}$ shifts both reflection eigenvalues equally and therefore does not affect the eigenstates.

At the ideal unperturbed EP, $\Delta r_{B}=\xi=0$ while $b\neq0$, and the two eigenstates coalesce into $|+\rangle$. An ideal nonreciprocal perturbation produces a finite $\Delta r_{B}$ while preserving $\xi=0$. The eigenvalue degeneracy is then lifted, but the reflection matrix remains triangular. Thus, the perturbed system is no longer at the EP, although $|+\rangle$ remains an exact eigenstate. The other eigenstate acquires a finite $|-\rangle$ component and moves away from the original EP polarization.

A realistic reflection matrix generally contains a small but nonzero $\xi$ owing to nonresonant background scattering, multiple reflections, or imperfect isolation of the two polarization channels. To determine how this residual conversion modifies the state-selective response, we write a right eigenstate as $|\psi\rangle\propto|+\rangle+\eta|-\rangle$. Substitution into the eigenvalue equation gives $b\eta^{2}+\Delta r_{B}\eta-\xi=0$, with the two solutions
\begin{equation}
  \eta_{\pm}=
  \frac{
    -\Delta r_{B}
    \pm
    \sqrt{\left(\Delta r_{B}\right)^{2}+4b\xi}
  }{2b}.
  \label{eq:eta}
\end{equation}
Equation~\eqref{eq:eta} shows that the eigenstate response is governed by the competition between the nonreciprocal channel splitting $\Delta r_{B}$ and the residual leakage $\xi$. When $|b\xi|\gtrsim|\Delta r_{B}|^{2}$, the residual channel substantially perturbs both branches. In the limiting case $\Delta r_{B}\rightarrow0$, the two eigenstates satisfy $\eta_{\pm}\simeq\pm\sqrt{\xi/b}$ and depart from the EP polarization with comparable amplitudes.

By contrast, when $|b\xi|\ll|\Delta r_{B}|^{2}$, the two solutions separate into parametrically different responses: $\eta_{\text{q-pin}}\simeq\xi/\Delta r_{B}$ and $\eta_{\text{tun}}\simeq-\Delta r_{B}/b-\xi/\Delta r_{B}$. The first branch contains only a small opposite-handed component and therefore remains quasi-pinned near the original EP polarization. The second branch is governed primarily by $-\Delta r_{B}/b$ and undergoes a substantially larger polarization evolution. Exact triangularity is therefore sufficient but not necessary for state-selective splitting; a finite reverse-conversion channel can be tolerated provided that it remains small compared with the induced channel splitting.

\begin{figure}[!htb]
\centering\includegraphics[width=8.5cm]{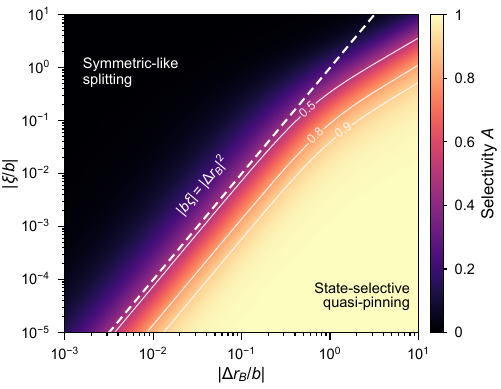}
\caption{
\textbf{Eigenstate selectivity near a second-order reflection-matrix EP.}
The color encodes the selectivity factor $A$ of Eq.~\eqref{eq:A}, where $\|\delta\mathbf{S}_{\rm q\text{-}pin}\|$ and $\|\delta\mathbf{S}_{\rm tun}\|$ are the deviation distances of the quasi-pinned and tunable eigenstates from the EP eigenstate, computed from Eq.~\eqref{eq:eta}. The axes are the normalized channel splitting $|\Delta r_B/b|$ and residual leakage $|\xi/b|$; thin contours mark $A=0.5$, $0.8$ and $0.9$. The dashed line, $|b\xi|=|\Delta r_B|^2$, marks the crossover between the leakage-dominated and quasi-pinning regimes. The map is evaluated for real, positive $b$, $\Delta r_B$ and $\xi$, i.e., for in-phase $\Delta r_B^2$ and $4b\xi$; in the quasi-pinning regime $|b\xi|\ll|\Delta r_B|^2$, $A$ is insensitive to this relative phase, whereas near and above the crossover the contours depend on it.}
\label{fig:phasemap}
\end{figure}

To visualize this crossover, we define the eigenstate-selectivity factor
\begin{equation}
  A=
  \frac{
    \lVert\delta\mathbf{S}_{\text{tun}}\rVert
    -
    \lVert\delta\mathbf{S}_{\text{q-pin}}\rVert
  }{
    \lVert\delta\mathbf{S}_{\text{tun}}\rVert
    +
    \lVert\delta\mathbf{S}_{\text{q-pin}}\rVert
  },
  \label{eq:A}
\end{equation}
where the two branches are identified by continuity from the ideal triangular limit. Figure~\ref{fig:phasemap} maps $A$ in the normalized parameter plane $\left(|\Delta r_{B}/b|,\,|\xi/b|\right)$. Values near $A=0$ indicate comparable eigenstate displacements, whereas $A\rightarrow1$ denotes a strongly state-selective response. The curve $|\xi/b|=|\Delta r_{B}/b|^{2}$ corresponds to $|b\xi|=|\Delta r_{B}|^{2}$ and marks the crossover between the leakage-dominated and quasi-pinning regimes. It should not be interpreted as a phase boundary, because the eigenstate response evolves continuously across this curve.%

The full-wave metasurface considered in the next section generally has $\xi\neq0$, so the exact pinning predicted by the ideal Hamiltonian is replaced by quasi-pinning in its reflection response. Nevertheless, the state-selective evolution remains robust when $|b\xi|\ll|\Delta r_{B}|^{2}$. We next identify the corresponding reflection-matrix EP in a magneto-optical metasurface and examine its evolution under a finite gyrotropic perturbation.

\section{State-selective eigenstate splitting in a magneto-optical metasurface}
\subsection{Metasurface design and identification of a reflection-matrix EP}
\label{sec:structure}
\begin{figure}[!htb]
\centering\includegraphics[width=\linewidth]{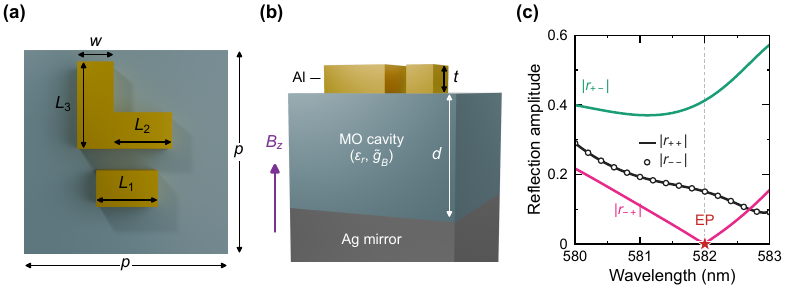}
\caption{
\textbf{Schematic of the metasurface structure and circular-polarization reflection spectra.}
\textbf{(a)} Top view of the unit cell, showing the upper aluminum-rod array; $w$, $L_1$--$L_3$ and $p$ denote the rod width, rod lengths and lattice period.
\textbf{(b)} Side view of the layer stack, consisting of the upper aluminum rods, middle magneto-optical cavity, and bottom silver mirror, with magneto-optical nonreciprocal coupling introduced by a perpendicular field $B_z$; $t$ and $d$ are the thicknesses of the aluminum rods and the cavity.
\textbf{(c)} Circular-polarization reflection amplitudes $|r_{++}|$ (black curve), $|r_{--}|$ (open circles), $|r_{+-}|$ (green curve) and $|r_{-+}|$ (magenta curve) of the metasurface over $580$--$583~\mathrm{nm}$; at $582~\mathrm{nm}$ (red star, vertical dashed line) the cross-polarized reflection coefficient $r_{-+}$ approaches zero, corresponding to the system reaching an EP.}
\label{fig:structure}
\end{figure}

Guided by the reflection-matrix model developed in Section~\ref{sec:robust}, we design a reflective magneto-optical metasurface that supports an LCP reflection-matrix EP in the reciprocal limit, $\tilde{g}_{B}=0$. As shown in Figs.~\ref{fig:structure}(a) and \ref{fig:structure}(b), the metasurface consists of an upper array of aluminum nanorods~\cite{rakic1995AlgorithmDeterminationIntrinsic}, each unit cell comprising an L-shaped rod and a detached straight rod, a substituted-iron-garnet magneto-optical cavity~\cite{zvezdin1997modern,onbasli2016optical,fakhrul2021high,hansen1983magnetic,yao2015faraday}, and a bottom silver mirror~\cite{johnson1972OpticalConstantsNoble}. The lattice period is $p=300$~nm. The aluminum nanorods have width $w=50$~nm, thickness $t=30$~nm, and lengths $L_{1}=75$~nm, $L_{2}=110$~nm, and $L_{3}=86$~nm. The magneto-optical cavity has a thickness of $d=2400$~nm and is described by the relative-permittivity tensor%
\begin{equation}
  \boldsymbol{\varepsilon}=
  \begin{pmatrix}
    \varepsilon_{r} & i\tilde{g}_{B} & 0\\[2pt]
    -i\tilde{g}_{B} & \varepsilon_{r} & 0\\[2pt]
    0 & 0 & \varepsilon_{r}
  \end{pmatrix},
  \qquad \varepsilon_{r}=5.5.
  \label{eq:eps}
\end{equation}
A magnetic bias normal to the metasurface plane generates the antisymmetric in-plane permittivity components. In the full-wave simulations, the resulting nonreciprocal perturbation is parameterized directly by $\tilde{g}_{B}$, whose sign reverses with the direction of the magnetic bias. Because $\varepsilon_{r}$ and $\tilde{g}_{B}$ are both real in this model, the tensor in Eq.~(\ref{eq:eps}) is Hermitian, $\boldsymbol{\varepsilon}=\boldsymbol{\varepsilon}^{\dagger}$, so the magneto-optical layer itself is lossless; the losses of the metasurface arise from the aluminum nanorods, the silver mirror, and radiation.

We first identify the unperturbed reflection-matrix EP at $\tilde{g}_{B}=0$. For each wavelength, the simulated co- and cross-polarized reflection coefficients are assembled into the circular-basis reflection matrix $R_{\mathrm{c}}$ introduced in Eq.~\eqref{eq:Rc_local}, where $r_{\alpha\beta}$ denotes reflection from incident polarization $\beta$ to outgoing polarization $\alpha$. In the reciprocal limit, the simulated co-polarized channels satisfy $r_{++}=r_{--}$ to within numerical precision. The reflection spectra in Fig.~\ref{fig:structure}(c) show that, near 582~nm, $r_{-+}$ is strongly suppressed while $r_{+-}$ remains finite. The reflection matrix therefore approaches the upper-triangular structure associated with the LCP EP discussed in Section~\ref{sec:robust}.

\begin{figure}[!htb]
\centering\includegraphics[width=\linewidth]{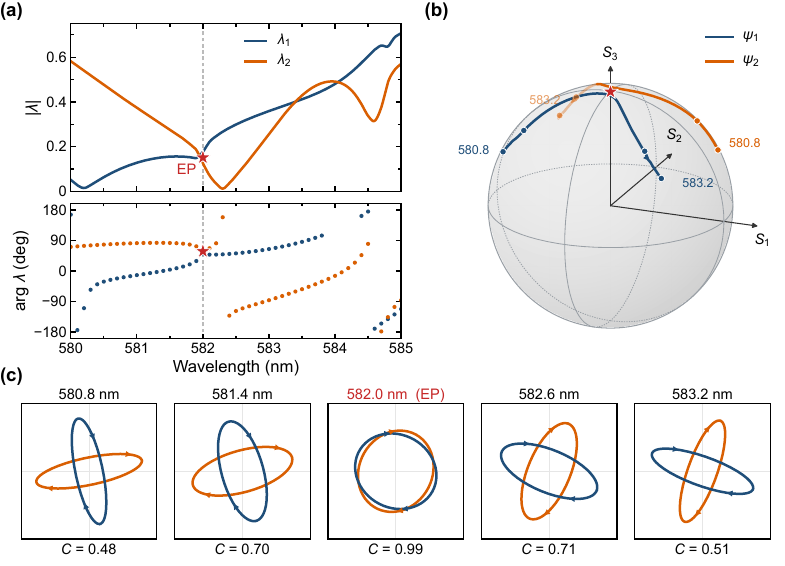}
\caption{
\textbf{Evolution of the reflection-matrix eigenvalues and eigenstates with wavelength and the determination of the EP.}
\textbf{(a)} Magnitudes $|\lambda|$ (upper panel, lines) and phases $\arg\lambda$ (lower panel, dots) of the two eigenvalues versus wavelength. At $582~\mathrm{nm}$ (red star, vertical dashed line) the magnitudes and phases of the two eigenvalues cross, marking the eigenvalue coalescence; $\lambda_1$ and $\lambda_2$ denote eigenvalue 1 and eigenvalue 2; the same color marks the same eigen-pair throughout (blue: $\lambda_1$, $|\psi_1\rangle$; orange: $\lambda_2$, $|\psi_2\rangle$).
\textbf{(b)} Trajectories of the normalized eigenstates $|\psi_1\rangle$ (blue) and $|\psi_2\rangle$ (orange) on the Poincar\'e sphere as the wavelength is swept from $580.8$ to $583.2~\mathrm{nm}$ (end points labeled); dots mark the five wavelengths shown in (c), and segments on the far hemisphere are drawn faded. At the EP the two eigenstates converge near the north pole (LCP), marked by a red star.
\textbf{(c)} Polarization ellipses of the two eigenstates (blue and orange) at five representative wavelengths ($580.8$, $581.4$, $582.0$, $582.6$, $583.2~\mathrm{nm}$). The overlap of the two polarization states is evaluated by $C=|\langle \psi_1 | \psi_2 \rangle|$. At the EP ($582~\mathrm{nm}$) the two eigenstates overlap and both approach circular polarization ($C=0.99$), while the overlap decreases away from the EP.}
\label{fig:eigenvalue}
\end{figure}

The suppression of a single conversion channel, however, is not by itself sufficient to establish an EP. We therefore examine both the eigenvalues and eigenstates of the full complex reflection matrix. Figure~\ref{fig:eigenvalue}(a) shows that the magnitudes and phases of the two reflection eigenvalues converge near $\lambda\simeq582$~nm, indicating complex-eigenvalue coalescence. Over the same wavelength interval, the two reflection eigenstates approach one another on the Poincar\'{e} sphere and converge near the LCP pole (Fig.~\ref{fig:eigenvalue}(b)). Their coalescence is shown more directly by the polarization ellipses in Fig.~\ref{fig:eigenvalue}(c). Defining the overlap between the normalized eigenstates as $C=|\langle\psi_{1}|\psi_{2}\rangle|$, we find that $C$ increases from 0.48 at 580.8~nm to 0.99 at the sampled wavelength closest to the EP, 582.0~nm, and decreases again to 0.51 at 583.2~nm. The simultaneous coalescence of the complex eigenvalues and polarization eigenstates therefore identifies a second-order LCP reflection-matrix EP near 582~nm. The expected square-root scaling of the eigenvalue splitting around this point is further verified in Supplementary Note~6.%

\subsection{State-selective reflection-eigenstate evolution under gyrotropic perturbation}
\label{sec:evolution}
Having identified the unperturbed reflection-matrix EP in Section~\ref{sec:structure}, we now examine how its two eigenstate branches evolve under a finite gyrotropic perturbation. The geometry and operating wavelength are held fixed at the sampled point closest to the EP, $\lambda=582.0~\mathrm{nm}$, while $\tilde{g}_{B}$ is varied. Thus, $\tilde{g}_{B}=0$ represents the unperturbed EP, whereas any finite $\tilde{g}_{B}$ lifts the eigenvalue degeneracy. At each value of $\tilde{g}_{B}$, the full complex reflection matrix is diagonalized, and the two branches are tracked continuously from their common state at $\tilde{g}_{B}=0$. We label the weakly displaced branch as State~1 and the tunable branch as State~2.

\begin{figure}[!htb]
\centering\includegraphics[width=\linewidth]{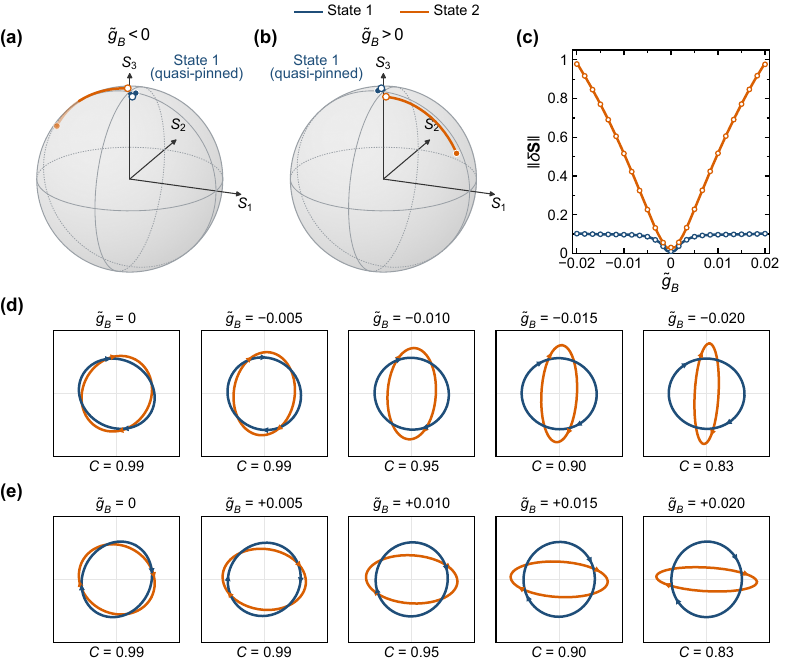}
\caption{
\textbf{Eigenstate evolution within $\tilde{g}_B \in [-0.02,\,+0.02]$.}
\textbf{(a)} Eigenstate trajectories on the Poincar\'e sphere for $\tilde{g}_B<0$; open circles mark the start points and filled circles the end points. State~1 is pinned near the north pole (LCP), while State~2 migrates down along the sphere.
\textbf{(b)} Eigenstate trajectories on the Poincar\'e sphere for $\tilde{g}_B>0$; open circles mark the start points and filled circles the end points. State~1 remains pinned at LCP, while the trajectory of State~2 is mirrored about the $S_3$ axis relative to the negative case.
\textbf{(c)} Euclidean distance $\|\delta\mathbf{S}\|$ of the two eigenstates from the EP versus $\tilde{g}_B$: State~1 (blue) remains small ($\lesssim 0.1$, quasi-pinned), while State~2 (orange) increases monotonically with $|\tilde{g}_B|$.
\textbf{(d), (e)} Eigen-polarization ellipses and overlap $C$ at five representative points for negative and positive $\tilde{g}_B$ ($\tilde{g}_B = 0,\,\pm0.005,\,\pm0.010,\,\pm0.015,\,\pm0.020$); $C$ drops from 0.99 (EP) to 0.83.}
\label{fig:gB_eigenstate_002}
\end{figure}

Figures~\ref{fig:gB_eigenstate_002}(a) and \ref{fig:gB_eigenstate_002}(b) show the resulting eigenstate trajectories on the Poincar\'e sphere for negative and positive $\tilde{g}_{B}$, respectively. For $\tilde{g}_{B}<0$, State~1 remains close to the original LCP EP polarization, whereas State~2 departs from LCP and evolves through elliptically polarized states along one side of the sphere (Fig.~\ref{fig:gB_eigenstate_002}(a)). Reversing the sign of $\tilde{g}_{B}$ leaves State~1 nearly unchanged but redirects State~2 along the opposite trajectory (Fig.~\ref{fig:gB_eigenstate_002}(b)). The two signs therefore do not represent different physical mechanisms: the magnitude $|\tilde{g}_{B}|$ controls the extent of the eigenstate displacement, while the sign of $\tilde{g}_{B}$ selects its direction on the Poincar\'e sphere.%

This branch-selective evolution is quantified in Fig.~\ref{fig:gB_eigenstate_002}(c) using the Stokes-vector displacement $\|\delta\mathbf{S}\|$ defined in Eq.~\eqref{eq:dS}. The displacement of State~1 remains small throughout the simulated range, whereas that of State~2 increases rapidly with $|\tilde{g}_{B}|$. Unlike the exactly pinned state of the ideal triangular Hamiltonian, State~1 exhibits a small but finite displacement in the full-wave calculation. This behavior is consistent with the quasi-pinning regime derived in Section~\ref{sec:robust}, in which a residual reverse-conversion channel weakly perturbs the nominally pinned branch while leaving the tunable branch predominantly controlled by the gyrotropically induced channel splitting.

The polarization ellipses in Figs.~\ref{fig:gB_eigenstate_002}(d) and \ref{fig:gB_eigenstate_002}(e) provide a direct representation of the same evolution. At $\tilde{g}_{B}=0$, the two eigenstates nearly coalesce, with an overlap $C=|\langle\psi_{1}|\psi_{2}\rangle|=0.99$. As $\tilde{g}_{B}$ is varied from 0 to $-0.02$, State~1 remains close to LCP, while State~2 develops an increasingly elliptical polarization, reducing the overlap to $C=0.83$. A comparable decrease occurs when $\tilde{g}_{B}$ is increased to $+0.02$, but the major-axis orientation of State~2 evolves in the opposite direction. The overlap $C$ quantifies the separation between the two eigenstates, while the individual values of $\|\delta\mathbf{S}\|$ distinguish the quasi-pinned and tunable branches.

The full-wave results therefore reproduce the central prediction of the reflection-matrix model. A gyrotropic perturbation applied to an initially tuned chiral EP produces unequal eigenstate evolution, with one reflection eigenstate remaining quasi-pinned near the EP polarization and the other being continuously steered by the magnitude and sign of $\tilde{g}_{B}$.

\subsection{Magnetically tunable reflection circular dichroism}
The state-selective eigenstate evolution discussed in Section~\ref{sec:evolution} is accompanied by a directly observable difference between the reflection responses to LCP and RCP incidence. This intensity-level response does not measure the reflection eigenstates themselves. Rather, it arises because the gyrotropic perturbation modifies the circular-polarization channels of the reflection matrix unequally.

\begin{figure}[!htb]
\centering\includegraphics[width=8.5cm]{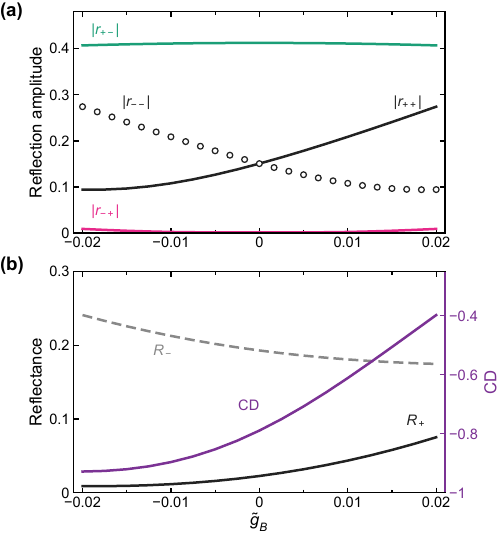}
\caption{
\textbf{Circular-polarization reflection magneto-optical tuning and circular dichroism within $\tilde{g}_B \in [-0.02,\,+0.02]$.}
\textbf{(a)} Reflection matrix elements $|r_{++}|$ (black line), $|r_{--}|$ (open circles), $|r_{+-}|$ (green) and $|r_{-+}|$ (magenta) versus $\tilde{g}_B$: $|r_{-+}|$ stays small over the whole range, $|r_{++}|$ and $|r_{--}|$ split with $\tilde{g}_B$, and $|r_{+-}|$ barely changes.
\textbf{(b)} LCP reflectance $R_+$ (black solid), RCP reflectance $R_-$ (gray dashed), and circular dichroism $\mathrm{CD}$ (purple, right axis) versus $\tilde{g}_B$; the CD evolves monotonically from $\sim-0.9$ to $\sim-0.4$.}
\label{fig:gB_CD_002}
\end{figure}

Figure~\ref{fig:gB_CD_002}(a) shows the magnitudes of the four circular-basis reflection coefficients as functions of $\tilde{g}_{B}$ at $\lambda=582.0~\mathrm{nm}$. Throughout the interval $\tilde{g}_{B}\in[-0.02,+0.02]$, the residual reverse-conversion coefficient $|r_{-+}|$ remains small, while $|r_{+-}|$ changes only weakly. By contrast, the co-polarized coefficients $|r_{++}|$ and $|r_{--}|$ exhibit a pronounced and oppositely directed variation. In particular, their responses satisfy the approximate relation $|r_{++}(\tilde{g}_{B})|\simeq|r_{--}(-\tilde{g}_{B})|$, showing that reversing the gyrotropic bias interchanges the dominant responses of the two circular input channels.%

For incident LCP and RCP waves, the total reflectances are respectively $R_{+}=|r_{++}|^{2}+|r_{-+}|^{2}$ and $R_{-}=|r_{--}|^{2}+|r_{+-}|^{2}$. We quantify their difference using the reflection circular dichroism
\begin{equation}
  \mathrm{CD}=
  \frac{R_{+}-R_{-}}{R_{+}+R_{-}}.
  \label{eq:CD}
\end{equation}
As shown in Fig.~\ref{fig:gB_CD_002}(b), the splitting of the co-polarized circular channels produces a continuous variation of both $R_{+}$ and $R_{-}$ and tunes the CD from approximately $-0.9$ to $-0.4$ over the simulated interval. The negative sign indicates stronger reflection for RCP than for LCP under the convention used here.

The CD is already finite at $\tilde{g}_{B}=0$ because the unperturbed chiral EP contains strongly asymmetric circular-polarization conversion, $r_{-+}\simeq0$ while $r_{+-}$ remains finite. Consequently, the CD response need not be antisymmetric about $\tilde{g}_{B}=0$. The gyrotropic perturbation acts on top of this structural scattering background by reversibly splitting $r_{++}$ and $r_{--}$, thereby providing continuous magnetic control of the reflection CD. An extended effective-parameter scan shows that the CD can also cross zero and reverse sign at larger $|\tilde{g}_{B}|$ (Supplementary Fig.~S2). This extended scan illustrates the complete tuning trend and should not be interpreted as the intrinsic gyrotropy of a specific garnet at 582~nm. In practice, a comparable effective perturbation may be obtained by increasing the magneto-optical interaction length, exploiting the double-pass geometry of the reflective configuration, introducing resonant cavity enhancement, or retuning the EP to a wavelength range with a larger magneto-optical figure of merit.

These results connect the eigenstate-level mechanism to a measurable polarization-selective response. The same nonreciprocal perturbation that leaves one reflection eigenstate quasi-pinned while steering the other also produces a reversible imbalance between the LCP and RCP reflectances. The metasurface therefore converts state-selective splitting near a chiral EP into actively tunable reflection circular dichroism.

\section{Conclusion}
We have theoretically identified and numerically demonstrated a state-selective eigenstate response induced by antisymmetric nonreciprocal coupling near a second-order exceptional point. In an ideal two-mode description, the nonreciprocal off-diagonal perturbation transforms the Hamiltonian into an upper-triangular form in the chiral basis, so that one eigenstate remains pinned to the EP polarization while the other is continuously reoriented by the perturbation. In realistic reflective metasurfaces, residual cross-polarized leakage relaxes exact pinning into quasi-pinning, but the branch-selective response persists when the magneto-optically induced circular-channel splitting dominates the residual leakage.

As a representative implementation, we designed a mirror-backed magneto-optical metasurface supporting a circular-polarization reflection-matrix EP near 582 nm. Full-wave simulations show that a perpendicular magnetic field drives the predicted state-selective response: one reflection eigenstate remains close to the EP circular-polarization state, whereas the other is reversibly displaced on the Poincaré sphere by the magnitude and sign of the magneto-optical coefficient. The same nonreciprocal perturbation splits the co-polarized circular reflection channels and converts the eigenstate-level asymmetry into a magnetically tunable reflection circular dichroism.

The present work focuses on the mechanism of nonreciprocal-perturbation-induced state-selective splitting, rather than the optimization of a particular magneto-optical material platform. These results suggest a nonreciprocal route to selective eigenstate control at EPs in two-mode non-Hermitian platforms supporting antisymmetric coupling, with potential relevance to active chiral metasurfaces, magnetically reconfigurable polarization devices, and time-reversal-broken photonic systems.

\section{Methods}
\subsection{Numerical simulations}
All electromagnetic simulations are based on the finite-element method (FEM), performed with the Wave Optics module of COMSOL Multiphysics. Floquet periodic boundary conditions are applied along $x$ and $y$ to model an infinite periodic array. Perfectly matched layers (PMLs) are placed above and below the simulation domain along $z$ to absorb outgoing waves and suppress numerical reflection.

\subsection{Material parameters}
The optical constants of the aluminum-rod array are taken from the Raki{\'c} 1995 model~\cite{rakic1995AlgorithmDeterminationIntrinsic}; those of the silver mirror are taken from the Johnson \& Christy 1972 experimental dataset~\cite{johnson1972OpticalConstantsNoble}. The magneto-optical cavity layer is modeled as a gyrotropic medium~\cite{zvezdin1997modern} whose relative permittivity tensor is given by Eq.~(\ref{eq:eps}), with $\varepsilon_r = 5.5$ and the dimensionless magneto-optical coefficient $\tilde{g}_B$ controlled by the applied perpendicular field $B_z$.
To keep the magneto-optical parameter consistent with the realistic magnitudes of common substituted iron-garnet materials~\cite{onbasli2016optical,fakhrul2021high,hansen1983magnetic,yao2015faraday}, the main-text simulations use $\tilde{g}_B \in [-0.02,\,+0.02]$. The extended sweep $|\tilde{g}_B| \leq 0.1$ is used only to display the full eigenstate-evolution trend, and can be understood as the equivalent nonreciprocal perturbation under strong effective gyrotropy, Bi-rich garnet resonant response, or cavity enhancement. Details of the gyrotropic model, the literature values of $\tilde{g}_B$ for substituted iron garnets, and the extended scan are given in Supplementary Note~2.
\subsection{Reflection matrix and eigenstate analysis}
For each wavelength and each value of $\tilde{g}_B$, $x$- and $y$-polarized plane waves are normally incident on the metasurface, and the simulated reflection S-parameters are used to construct the $2\times 2$ reflection Jones matrix $R$; a unitary transformation $R_c = U^\dagger R U$ (with $U=\frac{1}{\sqrt{2}}\bigl(\begin{smallmatrix}1 & 1 \\ i & -i\end{smallmatrix}\bigr)$) transforms it into the circular-polarization basis (LCP/RCP), yielding the four matrix elements $r_{++},r_{+-},r_{-+},r_{--}$ of $R_c$. Here, $|+\rangle$ ($|-\rangle$) denotes the laboratory-frame Jones vector $(1,\pm i)^{\mathsf T}/\sqrt{2}$, which corresponds to LCP (RCP) for the incident wave propagating along $-z$, with the $e^{i\omega t}$ convention and handedness defined facing the light. The reflected field is expressed in the same laboratory-fixed basis, so that the labels $\pm$ of the reflected components refer to this Jones basis rather than to the handedness of the counter-propagating reflected wave. The complex eigenvalues $\lambda_{1,2}$ and reflection eigenstates $|\psi_{1,2}\rangle$ are obtained by numerically diagonalizing $R_c$; the eigenstate trajectories on the Poincar\'e sphere are given by the corresponding normalized Stokes parameters $(S_1,S_2,S_3)$. The EP location is determined by simultaneously minimizing the eigenvalue difference $|\lambda_1-\lambda_2|$ and maximizing the eigenstate overlap $C=|\langle\psi_1|\psi_2\rangle|$ in the parameter space spanned by wavelength and $\tilde{g}_B$.
\subsection{Stokes parameters}
The Stokes parameters used to represent the eigenstates on the Poincar\'e sphere are defined as
\begin{equation}
\begin{aligned}
s_0 &= |E_x|^2+|E_y|^2, \\
s_1 &= |E_x|^2-|E_y|^2, \\
s_2 &= 2\operatorname{Re}(E_x^*E_y), \\
s_3 &= 2\operatorname{Im}(E_x^*E_y).
\end{aligned}
\end{equation}
The normalized Stokes parameters are
\begin{equation}
S_1=\frac{s_1}{s_0},
\qquad
S_2=\frac{s_2}{s_0},
\qquad
S_3=\frac{s_3}{s_0}.
\end{equation}

\begin{backmatter}
\bmsection{Funding}
National Natural Science Foundation of China (NSFC) (Grant No.~12304417); Natural Science Foundation of Heilongjiang Province of China (Grant No.~YQ2024A007); The Fundamental Research Funds for the Central Universities (Grant No.~XNJKKGYDJ2026019) and the National Research Foundation of Korea (NRF) through the government of Korea (NRF-RS-2026-25491025, NRF-RS-2026-25506975).

\bmsection{Disclosures}
The authors declare no conflicts of interest.

\bmsection{Data availability}
Data underlying the results presented in this paper are not publicly available at this time but may be obtained from the authors upon reasonable request.

\bmsection{Supporting information}
See Supplementary material for supporting content. Supplementary Note~1 derives the temporal coupled-mode model with nonreciprocal coupling and the resulting circular-basis reflection matrix; Supplementary Note~2 describes the gyrotropic permittivity model, estimates realistic values of $\tilde{g}_B$, and presents the extended $|\tilde{g}_B|\le0.1$ scan and a thin-cavity control simulation; Supplementary Note~3 gives the quasi-pinning analysis; Supplementary Note~4 reports the coupled-mode fit of the simulated reflection matrix; Supplementary Note~5 shows the magnetic-field-induced non-circular EPs; and Supplementary Note~6 verifies the square-root scaling of the eigenvalue splitting near the EP.
\end{backmatter}

\bibliography{references}

\begin{thebibliography}{10}
\newcommand{\enquote}[1]{``#1''}

\bibitem{elganainy2018NonHermitianPhysics}
R.~El-Ganainy, K.~G. Makris, M.~Khajavikhan, \emph{et~al.},
  \enquote{{Non-Hermitian physics and PT symmetry},}
  {\protect\JournalTitle{{Nature Physics}}} \textbf{14}, 11--19 (2018).

\bibitem{miri2019ExceptionalPointsOptics}
M.-A. Miri and A.~Al{\`u}, \enquote{{Exceptional points in optics and
  photonics},} {\protect\JournalTitle{{Science}}} \textbf{363}, eaar7709
  (2019).

\bibitem{ozdemir2019ParitytimeSymmetryExceptional}
{\c S}.~K. {\"O}zdemir, S.~Rotter, F.~Nori, and L.~Yang, \enquote{{Parity--time
  symmetry and exceptional points in photonics},}
  {\protect\JournalTitle{{Nature Materials}}} \textbf{18}, 783--798 (2019).

\bibitem{meng2024exceptional}
H.~Meng, Y.~S. Ang, and C.~H. Lee, \enquote{{Exceptional points in
  non-Hermitian systems: Applications and recent developments},}
  {\protect\JournalTitle{Applied Physics Letters}} \textbf{124}, 060502 (2024).

\bibitem{heiss2012PhysicsExceptionalPoints}
W.~D. Heiss, \enquote{{The physics of exceptional points},}
  {\protect\JournalTitle{{Journal of Physics A: Mathematical and Theoretical}}}
  \textbf{45}, 444016 (2012).

\bibitem{berry2004PhysicsNonhermitianDegeneracies}
M.~Berry, \enquote{{Physics of Nonhermitian Degeneracies},}
  {\protect\JournalTitle{{Czechoslovak Journal of Physics}}} \textbf{54},
  1039--1047 (2004).

\bibitem{qin2026PhotonicExceptionalPoints}
H.~Qin, W.~Lv, Z.~Zhang, \emph{et~al.}, \enquote{{Photonic exceptional points
  in engineered materials and their emerging applications},}
  {\protect\JournalTitle{{Nature Reviews Materials}}} \textbf{11}, 453--468
  (2026).

\bibitem{lee2023chiral}
H.~Lee, A.~Kecebas, F.~Wang, \emph{et~al.}, \enquote{{Chiral exceptional point
  and coherent suppression of backscattering in silicon microring with low loss
  Mie scatterer},} {\protect\JournalTitle{{eLight}}} \textbf{3}, 20 (2023).

\bibitem{chen2017ExceptionalPointsEnhance}
W.~Chen, {\c S}.~Kaya~{\"O}zdemir, G.~Zhao, \emph{et~al.},
  \enquote{{Exceptional points enhance sensing in an optical microcavity},}
  {\protect\JournalTitle{{Nature}}} \textbf{548}, 192--196 (2017).

\bibitem{hodaei2017EnhancedSensitivityHigher}
H.~Hodaei, A.~U. Hassan, S.~Wittek, \emph{et~al.}, \enquote{{Enhanced
  sensitivity at higher-order exceptional points},}
  {\protect\JournalTitle{{Nature}}} \textbf{548}, 187--191 (2017).

\bibitem{brandstetter2014ReversingPumpDependence}
M.~Brandstetter, M.~Liertzer, C.~Deutsch, \emph{et~al.}, \enquote{{Reversing
  the pump dependence of a laser at an exceptional point},}
  {\protect\JournalTitle{{Nature Communications}}} \textbf{5}, 4034 (2014).

\bibitem{doppler2016DynamicallyEncirclingExceptional}
J.~Doppler, A.~A. Mailybaev, J.~B{\"o}hm, \emph{et~al.}, \enquote{{Dynamically
  encircling an exceptional point for asymmetric mode switching},}
  {\protect\JournalTitle{{Nature}}} \textbf{537}, 76--79 (2016).

\bibitem{nasari2022ObservationChiralState}
H.~Nasari, G.~Lopez-Galmiche, H.~E. Lopez-Aviles, \emph{et~al.},
  \enquote{{Observation of chiral state transfer without encircling an
  exceptional point},} {\protect\JournalTitle{{Nature}}} \textbf{605}, 256--261
  (2022).

\bibitem{yoon2018TimeAsymmetricLoop}
J.~W. Yoon, Y.~Choi, C.~Hahn, \emph{et~al.}, \enquote{{Time-asymmetric loop
  around an exceptional point over the full optical communications band},}
  {\protect\JournalTitle{{Nature}}} \textbf{562}, 86--90 (2018).

\bibitem{yu2014FlatOpticsDesigner}
N.~Yu and F.~Capasso, \enquote{{Flat optics with designer metasurfaces},}
  {\protect\JournalTitle{{Nature Materials}}} \textbf{13}, 139--150 (2014).

\bibitem{yu2011LightPropagationPhase}
N.~Yu, P.~Genevet, M.~A. Kats, \emph{et~al.}, \enquote{{Light Propagation with
  Phase Discontinuities: Generalized Laws of Reflection and Refraction},}
  {\protect\JournalTitle{{Science}}} \textbf{334}, 333--337 (2011).

\bibitem{kildishev2013PlanarPhotonicsMetasurfaces}
A.~V. Kildishev, A.~Boltasseva, and V.~M. Shalaev, \enquote{{Planar Photonics
  with Metasurfaces},} {\protect\JournalTitle{{Science}}} \textbf{339}, 1232009
  (2013).

\bibitem{kodigala2016ExceptionalPointsThree}
A.~Kodigala, T.~Lepetit, and B.~Kant{\'e}, \enquote{{Exceptional points in
  three-dimensional plasmonic nanostructures},}
  {\protect\JournalTitle{{Physical Review B}}} \textbf{94}, 201103 (2016).

\bibitem{song2021PlasmonicTopologicalMetasurface}
Q.~Song, M.~Odeh, J.~Zuniga-Perez, \emph{et~al.}, \enquote{{Plasmonic
  topological metasurface by encircling an exceptional point},}
  {\protect\JournalTitle{{Science}}} \textbf{373}, 1133--1137 (2021).

\bibitem{he2023ScatteringExceptionalPoint}
T.~He, Z.~Zhang, J.~Zhu, \emph{et~al.}, \enquote{{Scattering exceptional point
  in the visible},} {\protect\JournalTitle{Light Sci. Appl.}} \textbf{12}, 229
  (2023).

\bibitem{shou2025ResonantScatteringExceptional}
Y.~Shou, D.~Wang, Y.~Wang, \emph{et~al.}, \enquote{{Resonant and scattering
  exceptional points in non-Hermitian metasurfaces},}
  {\protect\JournalTitle{{npj Nanophotonics}}} \textbf{2}, 29 (2025).

\bibitem{dong2020LossAssistedMetasurface}
S.~Dong, G.~Hu, Q.~Wang, \emph{et~al.}, \enquote{{Loss-Assisted Metasurface at
  an Exceptional Point},} {\protect\JournalTitle{{ACS Photonics}}} \textbf{7},
  3321--3327 (2020).

\bibitem{zhen2015SpawningRingsExceptional}
B.~Zhen, C.~W. Hsu, Y.~Igarashi, \emph{et~al.}, \enquote{{Spawning rings of
  exceptional points out of Dirac cones},} {\protect\JournalTitle{{Nature}}}
  \textbf{525}, 354--358 (2015).

\bibitem{sweeney2019PerfectlyAbsorbingExceptional}
W.~R. Sweeney, C.~W. Hsu, S.~Rotter, and A.~D. Stone, \enquote{{Perfectly
  Absorbing Exceptional Points and Chiral Absorbers},}
  {\protect\JournalTitle{{Physical Review Letters}}} \textbf{122}, 093901
  (2019).

\bibitem{park2020ObservationExceptionalPoint}
S.~H. Park, S.-G. Lee, S.~Baek, \emph{et~al.}, \enquote{{Observation of an
  exceptional point in a non-Hermitian metasurface},}
  {\protect\JournalTitle{{Nanophotonics}}} \textbf{9}, 1031--1039 (2020).

\bibitem{kang2016ChiralExceptionalPoints}
M.~Kang, J.~Chen, and Y.~D. Chong, \enquote{{Chiral exceptional points in
  metasurfaces},} {\protect\JournalTitle{{Physical Review A}}} \textbf{94},
  033834 (2016).

\bibitem{yang2024CreatingPairsExceptional}
Z.~Yang, P.-S. Huang, Y.-T. Lin, \emph{et~al.}, \enquote{{Creating pairs of
  exceptional points for arbitrary polarization control: asymmetric vectorial
  wavefront modulation},} {\protect\JournalTitle{{Nature Communications}}}
  \textbf{15}, 232 (2024).

\bibitem{qin2025SphereArbitrarilyPolarized}
H.~Qin, Z.~Yang, P.-S. Huang, \emph{et~al.}, \enquote{{Sphere of arbitrarily
  polarized exceptional points with a single planar metasurface},}
  {\protect\JournalTitle{{Nature Communications}}} \textbf{16}, 2656 (2025).

\bibitem{wu2024TwinsExceptionalPoints}
X.~Wu, X.~Zhao, Y.~Lin, \emph{et~al.}, \enquote{{Twins of Exceptional Points
  with Opposite Chirality for Non-Hermitian Metasurfaces},}
  {\protect\JournalTitle{{ACS Photonics}}} \textbf{11}, 2054--2060 (2024).

\bibitem{lawrence2014ManifestationPT}
M.~Lawrence, N.~Xu, X.~Zhang, \emph{et~al.}, \enquote{{Manifestation of P T
  Symmetry Breaking in Polarization Space with Terahertz Metasurfaces},}
  {\protect\JournalTitle{{Physical Review Letters}}} \textbf{113}, 093901
  (2014).

\bibitem{ergoktas2022TopologicalEngineeringTerahertz}
M.~S. Ergoktas, S.~Soleymani, N.~Kakenov, \emph{et~al.}, \enquote{{Topological
  engineering of terahertz light using electrically tunable exceptional point
  singularities},} {\protect\JournalTitle{{Science}}} \textbf{376}, 184--188
  (2022).

\bibitem{baek2023NonHermitianChiral}
S.~Baek, S.~H. Park, D.~Oh, \emph{et~al.}, \enquote{{Non-Hermitian chiral
  degeneracy of gated graphene metasurfaces},} {\protect\JournalTitle{Light
  Sci. Appl.}} \textbf{12}, 87 (2023).

\bibitem{ding2024ElectricallyTunableTopological}
F.~Ding, Y.~Deng, C.~Meng, \emph{et~al.}, \enquote{{Electrically tunable
  topological phase transition in non-Hermitian optical MEMS metasurfaces},}
  {\protect\JournalTitle{{Science Advances}}} \textbf{10}, eadl4661 (2024).

\bibitem{wang2025PhotoswitchableExceptionalPoints}
L.~Wang, H.~Liu, J.~Liu, \emph{et~al.}, \enquote{{Photoswitchable exceptional
  points derived from bound states in the continuum},}
  {\protect\JournalTitle{Light Sci. Appl.}} \textbf{14}, 377 (2025).

\bibitem{li2026exceptional}
Z.~Li, R.~Cai, X.~Wang, \emph{et~al.}, \enquote{{Exceptional deficiency of
  non-Hermitian systems},} {\protect\JournalTitle{Nature Physics}} \textbf{22},
  962--970 (2026).

\bibitem{ryu2024ExceptionalClassifications}
J.-W. Ryu, J.-H. Han, C.-H. Yi, \emph{et~al.}, \enquote{{Exceptional
  classifications of non-Hermitian systems},} {\protect\JournalTitle{{Commun.
  Phys.}}} \textbf{7}, 109 (2024).

\bibitem{sounas2017NonReciprocalPhotonics}
D.~L. Sounas and A.~Al{\`u}, \enquote{{Non-reciprocal photonics based on time
  modulation},} {\protect\JournalTitle{{Nature Photonics}}} \textbf{11},
  774--783 (2017).

\bibitem{bi2011ChipOpticalIsolation}
L.~Bi, J.~Hu, P.~Jiang, \emph{et~al.}, \enquote{{On-chip optical isolation in
  monolithically integrated non-reciprocal optical resonators},}
  {\protect\JournalTitle{{Nature Photonics}}} \textbf{5}, 758--762 (2011).

\bibitem{khanikaev2010OneWayExtraordinary}
A.~B. Khanikaev, S.~H. Mousavi, G.~Shvets, and Y.~S. Kivshar, \enquote{{One-Way
  Extraordinary Optical Transmission and Nonreciprocal Spoof Plasmons},}
  {\protect\JournalTitle{{Physical Review Letters}}} \textbf{105}, 126804
  (2010).

\bibitem{fan2019MagnetoopticalMetamaterialsNonreciprocal}
B.~Fan, M.~E. Nasir, L.~H. Nicholls, \emph{et~al.}, \enquote{{Magneto-Optical
  Metamaterials: Nonreciprocal Transmission and Faraday Effect Enhancement},}
  {\protect\JournalTitle{{Advanced Optical Materials}}} \textbf{7}, 1801420
  (2019).

\bibitem{belotelov2011EnhancedMagnetoOptical}
V.~I. Belotelov, I.~A. Akimov, M.~Pohl, \emph{et~al.}, \enquote{{Enhanced
  magneto-optical effects in magnetoplasmonic crystals},}
  {\protect\JournalTitle{{Nature Nanotechnology}}} \textbf{6}, 370--376 (2011).

\bibitem{belotelov2013PlasmonMediatedMagneto}
V.~I. Belotelov, L.~E. Kreilkamp, I.~A. Akimov, \emph{et~al.},
  \enquote{{Plasmon-mediated magneto-optical transparency},}
  {\protect\JournalTitle{{Nature Communications}}} \textbf{4}, 2128 (2013).

\bibitem{fan2003TemporalCoupledMode}
S.~Fan, W.~Suh, and J.~D. Joannopoulos, \enquote{{Temporal coupled-mode theory
  for the Fano resonance in optical resonators},} {\protect\JournalTitle{J.
  Opt. Soc. Am. A}} \textbf{20}, 569 (2003).

\bibitem{rakic1995AlgorithmDeterminationIntrinsic}
A.~D. Raki{\'c}, \enquote{{Algorithm for the determination of intrinsic optical
  constants of metal films: application to aluminum},}
  {\protect\JournalTitle{{Applied Optics}}} \textbf{34}, 4755--4767 (1995).

\bibitem{zvezdin1997modern}
A.~Zvezdin and V.~Kotov, \emph{Modern magnetooptics and magnetooptical
  materials} (CRC Press, 1997).

\bibitem{onbasli2016optical}
M.~C. Onbasli, L.~Beran, M.~Zahradn{\'\i}k, \emph{et~al.}, \enquote{{Optical
  and magneto-optical behavior of cerium yttrium iron garnet thin films at
  wavelengths of 200--1770 nm},} {\protect\JournalTitle{Sci. Rep.}} \textbf{6},
  23640 (2016).

\bibitem{fakhrul2021high}
T.~Fakhrul, S.~Tazlaru, B.~Khurana, \emph{et~al.}, \enquote{{High figure of
  merit magneto-optical Ce- and Bi-substituted terbium iron garnet films
  integrated on Si},} {\protect\JournalTitle{Advanced Optical Materials}}
  \textbf{9}, 2100512 (2021).

\bibitem{hansen1983magnetic}
P.~Hansen, K.~Witter, and W.~Tolksdorf, \enquote{{Magnetic and magneto-optic
  properties of lead- and bismuth-substituted yttrium iron garnet films},}
  {\protect\JournalTitle{Physical Review B}} \textbf{27}, 6608--6625 (1983).

\bibitem{yao2015faraday}
S.~Yao, T.~Sato, K.~Kaneko, \emph{et~al.}, \enquote{{Faraday effect of bismuth
  iron garnet thin film prepared by mist CVD method},}
  {\protect\JournalTitle{Japanese Journal of Applied Physics}} \textbf{54},
  063001 (2015).

\bibitem{johnson1972OpticalConstantsNoble}
P.~B. Johnson and R.~W. Christy, \enquote{{Optical constants of the noble
  metals},} {\protect\JournalTitle{{Physical Review B}}} \textbf{6}, 4370--4379
  (1972).

\end{thebibliography}

\end{document}